\documentclass[11pt]{article}
\usepackage{amsmath,epsfig,layout}
\begin{document}

\font\sixrm=cmr6

% Definitions and abbreviations

% Roman letters in math formulae

\def\rmd{{\rm d}}
\def\rmD{{\rm D}}
\def\rme{{\rm e}}
\def\rmO{{\rm O}}

% bold face letters in math formulae

\def\bfp{{\bf p}}
\def\bfq{{\bf q}}
\def\bfr{{\bf r}}
\def\bfs{{\bf s}}
\def\bft{{\bf t}}
\def\bfu{{\bf u}}
\def\bfv{{\bf v}}
\def\bfw{{\bf w}}
\def\bfx{{\bf x}}
\def\bfy{{\bf y}}
\def\bfz{{\bf z}}

% Real and integer numbers

\def\rz{\blackboardrrm}
\def\gz{\blackboardzrm}
\def\Im{{\rm Im}\,}
\def\Re{{\rm Re}\,}

% Special relations and symbols

\def\defeq{\mathrel{\mathop=^{\rm def}}}
\def\proof{\noindent{\sl Proof:}\kern0.6em}
\def\endproof{\hskip0.6em plus0.1em minus0.1em
\setbox0=\null\ht0=5.4pt\dp0=1pt\wd0=5.3pt
\vbox{\hrule height0.8pt
\hbox{\vrule width0.8pt\box0\vrule width0.8pt}
\hrule height0.8pt}}
\def\frac#1#2{\hbox{$#1\over#2$}}
\def\dual{\mathstrut^*\kern-0.1em}
\def\mod{\;\hbox{\rm mod}\;}
\def\ring{\mathaccent"7017}
\def\lvec#1{\setbox0=\hbox{$#1$}
    \setbox1=\hbox{$\scriptstyle\leftarrow$}
    #1\kern-\wd0\smash{
    \raise\ht0\hbox{$\raise1pt\hbox{$\scriptstyle\leftarrow$}$}}
    \kern-\wd1\kern\wd0}
\def\rvec#1{\setbox0=\hbox{$#1$}
    \setbox1=\hbox{$\scriptstyle\rightarrow$}
    #1\kern-\wd0\smash{
    \raise\ht0\hbox{$\raise1pt\hbox{$\scriptstyle\rightarrow$}$}}
    \kern-\wd1\kern\wd0}

% Lattice derivatives

\def\nab#1{{\nabla_{#1}}}
\def\nabstar#1{\nabla\kern-0.5pt\smash{\raise 4.5pt\hbox{$\ast$}}
               \kern-4.5pt_{#1}}
\def\drv#1{{\partial_{#1}}}
\def\drvstar#1{\partial\kern-0.5pt\smash{\raise 4.5pt\hbox{$\ast$}}
               \kern-5.0pt_{#1}}

% Lattice momenta

\def\momp#1#2{
    \setbox0=\hbox{${#1}$}\setbox1=\hbox{${#1}_{#2}$}
    {#1}_{#2}\kern-\wd1\kern\wd0
    \smash{\raise4.5pt\hbox{$\scriptscriptstyle +$}}}
\def\momm#1#2{
    \setbox0=\hbox{${#1}$}\setbox1=\hbox{${#1}_{#2}$}
    {#1}_{#2}\kern-\wd1\kern\wd0
    \smash{\raise4.5pt\hbox{$\scriptscriptstyle -$}}}
\def\mompm#1#2{
    \setbox0=\hbox{${#1}$}\setbox1=\hbox{${#1}_{#2}$}
    {#1}_{#2}\kern-\wd1\kern\wd0
    \smash{\raise4.5pt\hbox{$\scriptscriptstyle \pm$}}}
\def\smomp#1#2{
    \setbox0=\hbox{${#1}$}\setbox1=\hbox{${#1}_{#2}$}
    {#1}_{#2}\kern-\wd1\kern\wd0
    \smash{\raise3pt\hbox{$\scriptscriptstyle +$}}}
\def\smomm#1#2{
    \setbox0=\hbox{${#1}$}\setbox1=\hbox{${#1}_{#2}$}
    {#1}_{#2}\kern-\wd1\kern\wd0
    \smash{\raise3pt\hbox{$\scriptscriptstyle -$}}}
\def\smompm#1#2{
    \setbox0=\hbox{${#1}$}\setbox1=\hbox{${#1}_{#2}$}
    {#1}_{#2}\kern-\wd1\kern\wd0
    \smash{\raise3pt\hbox{$\scriptscriptstyle \pm$}}}
\def\si{\kern1pt{\rm si}}
\def\co{\kern1pt{\rm co}}

% Units

\def\MeV{{\rm MeV}}
\def\GeV{{\rm GeV}}
\def\TeV{{\rm TeV}}
\def\fm{{\rm fm}}

\def\fpi{F_{\pi}}

% Constants

\def\euler{\gamma_{\rm E}}

% Fields

\def\Nf{N_{\rm f}}
\def\psibar{\bar{\psi}}
\def\psiL{\psi_{\rm L}}
\def\psiR{\psi_{\rm R}}
\def\psibarL{\psibar_{\rm L}}
\def\psibarR{\psibar_{\rm R}}
\def\psiclass{\psi_{\rm cl}}
\def\psibarclass{\psibar_{\rm cl}}
\def\psitilde{\widetilde{\psi}}
\def\psiprime{\psi\kern1pt'}
\def\psibarprime{\psibar\kern1pt'}
\def\rhoprime{\rho\kern1pt'}
\def\rhobar{\bar{\rho}}
\def\rhobarprime{\rhobar\kern1pt'}
\def\rhobartilde{\kern2pt\tilde{\kern-2pt\rhobar}}
\def\rhobartildeprime{\kern2pt\tilde{\kern-2pt\rhobar}\kern1pt'}
\def\etabar{\bar{\eta}}
\def\chibar{\overline{\chi}}
\def\phibar{\overline{\phi}}
\def\zetabar{\bar{\zeta}}
\def\zetaprime{\zeta\kern1pt'}
\def\zetabarprime{\zetabar\kern1pt'}
\def\zetar{\zeta_{\raise-1pt\hbox{\sixrm R}}}
\def\zetabarr{\zetabar_{\raise-1pt\hbox{\sixrm R}}}
\def\phieff{\phi_{\rm eff}}
\def\phiimpr{\phi_{\kern0.5pt\hbox{\sixrm I}}}
\def\phir{\phi_{\hbox{\sixrm R}}}
\def\ar{A_{\hbox{\sixrm R}}}
\def\vr{V_{\hbox{\sixrm R}}}
\def\pr{P_{\hbox{\sixrm R}}}
\def\sr{S_{\hbox{\sixrm R}}}
\def\aimp{A_{\hbox{\sixrm I}}}
\def\vimp{V_{\hbox{\sixrm I}}}
\def\pimp{P_{\hbox{\sixrm I}}}
\def\simp{S_{\hbox{\sixrm I}}}
\def\ce{{\cal E}}

% Dirac matrices

\def\dirac#1{\gamma_{#1}}
\def\diracstar#1#2{
    \setbox0=\hbox{$\gamma$}\setbox1=\hbox{$\gamma_{#1}$}
    \gamma_{#1}\kern-\wd1\kern\wd0
    \smash{\raise4.5pt\hbox{$\scriptstyle#2$}}}

\def\Pm{P_-}
\def\Pp{P_+}

\def\Rap{R_{\rm AP}}
\def\Rm{R_{\rm m}}
\def\Rz{R_{\rm Z}}
\def\Cf{C_{\rm F}}

% Improvement coefficients

\def\bx{b_{\rm X}}
\def\ba{b_{\rm A}}
\def\tba{\tilde{b}_{\rm A}}
\def\bp{b_{\rm P}}
\def\bs{b_{\rm S}}
\def\bv{b_{\rm V}}
\def\tbv{\tilde{b}_{\rm V}}
\def\bt{b_{\rm T}}
\def\bap{b_{\rm A,P}}
\def\bg{b_{\rm g}}
\def\bm{b_{\rm m}}
\def\tbm{\tilde{b}_{\rm m}}
\def\bmu{b_{\mu}}
\def\bzeta{b_{\zeta}}
\def\bmSF{b_{\rm m}^{\hbox{\sixrm SF}}}

\def\cx{c_{\rm X}}
\def\ca{c_{\rm A}}
\def\cv{c_{\rm V}}
\def\Ca{\ca}
\def\Cv{\cv}
\def\csw{c_{\rm sw}}
\def\cs{c_{\rm s}}
\def\ct{c_{\rm t}}
\def\cst{c_{\rm s,t}}
\def\ctildes{\tilde{c}_{\rm s}}
\def\ctildet{\tilde{c}_{\rm t}}
\def\ctildest{\tilde{c}_{\rm s,t}}

% Correlation functions

\def\fa{f_{\rm A}}
\def\fda{f_{\delta{\rm A}}}
\def\fp{f_{\rm P}}
\def\fx{f_{\rm X}}
\def\kv{k_{\rm V}}
\def\fv{f_{\rm V}}
\def\kt{k_{\rm T}}
\def\f1{f_1}
\def\hx{h_{\rm X}}
\def\ha{h_{\rm A}}
\def\hda{h_{\rm dA}}
\def\hdv{h_{\rm dV}}
\def\hp{h_{\rm P}}
\def\hv{h_{\rm V}}
\def\h1{h_1}

% Gauge group

\def\SUtwo{{\rm SU(2)}}
\def\SUthree{{\rm SU(3)}}
\def\SUn{{\rm SU}(N)}
\def\tr{\,\hbox{tr}\,}
\def\Ad{{\rm Ad}\,}
\def\CF{C_{\rm F}}
\def\cf{\CF}

% Action

\def\Sg{S_{\rm G}}
\def\Sf{S_{\rm F}}
\def\Seff{S_{\rm eff}}
\def\Simpr{S_{\rm impr}}
\def\Simprf{S_{\rm F\!,impr}}
\def\Zf{{\cal Z}_{\rm F}}
\def\op#1{{\cal O}_{\rm #1}}
\def\opprime#1{\setbox0=\hbox{${\cal O}$}\setbox1=\hbox{${\cal O}_{\rm #1}$}
    {\cal O}_{\rm #1}\kern-\wd1\kern\wd0
    \smash{\raise4.5pt\hbox{\kern1pt$\scriptstyle\prime$}}\kern1pt}
\def\ophat#1{\widehat{\cal O}_{\rm #1}}
\def\ophatprime#1{\setbox0=\hbox{$\widehat{\cal O}$}
    \setbox1=\hbox{$\widehat{\cal O}_{\rm #1}$}
    \widehat{\cal O}_{\rm #1}\kern-\wd1\kern\wd0
    \smash{\raise4.5pt\hbox{\kern1pt$\scriptstyle\prime$}}\kern1pt}
\def\bop#1{{\cal L}_{\rm #1}}
\def\bopprime#1{\setbox0=\hbox{${\cal O}$}\setbox1=\hbox{${\cal O}_{\rm #1}$}
    {\cal L}_{\rm #1}\kern-\wd1\kern\wd0
    \smash{\raise4.5pt\hbox{\kern1pt$\scriptstyle\prime$}}\kern1pt}
\def\blag#1{{\cal B}_{#1}}
\def\blagprime#1{\setbox0=\hbox{${\cal B}$}\setbox1=\hbox{${\cal B}_{#1}$}
    {\cal B}_{#1}\kern-\wd1\kern\wd0
    \smash{\raise5.2pt\hbox{\kern1pt$\scriptstyle\prime$}}\kern1pt}

% Renormalization schemes

\def\alphaSF{\alpha_{\rm SF}}
\def\alphaTP{\alpha_{\rm TP}}
\def\alphaMSbar{\alpha_{\msbar}}

\def\gms{g_{\ms}}
\def\gbar{\bar{g}}
\def\gbarMS{\gbar_{\ms}}
\def\gbarMSbar{\gbar_{\msbar}}
\def\gbarSF{\gbar_{\rm SF}}
\def\gbarTP{\gbar_{\rm TP}}
\def\gr{g_{{\hbox{\sixrm R}}}}
\def\gR{\gr}
\def\glat{g_{\lat}}
\def\gSF{g_{{\hbox{\sixrm SF}}}}
\def\gp{g_{\rm P}}

\def\muq{\mu_{\rm q}}
\def\mq{m_{\rm q}}
\def\mqtilde{\widetilde{m}_{\rm q}}
\def\muqtilde{\widetilde{\mu}_{\rm q}}

\def\mr{m_{{\hbox{\sixrm R}}}}
\def\mur{\mu_{{\hbox{\sixrm R}}}}
\def\mc{m_{\rm c}}
\def\mpole{m_{\rm p}}
\def\mlat{m_{\lat}}
\def\mSF{m_{\hbox{\sixrm SF}}}

\def\zx{Z_{\rm X}}
\def\za{Z_{\rm A}}
\def\zv{Z_{\rm V}}
\def\zp{Z_{\rm P}}
\def\zs{Z_{\rm S}}
\def\zg{Z_{\rm g}}
\def\zm{Z_{\rm m}}
\def\zmu{Z_{\mu}}
\def\zgm{Z_{\rm g,m}}
\def\zphi{Z_{\phi}}
\def\zmSF{Z_{\rm m}^{\hbox{\sixrm SF}}}
\def\zmlat{Z_{\rm m}^{\rm lat}}
\def\zzeta{Z_{\zeta}}

\def\Zx{\zx}
\def\Za{\za}
\def\Zv{\zv}
\def\Zp{\zp}
\def\Zs{\zs}
\def\Zg{\zg}
\def\Zm{\zm}
\def\Zmu{Z_{\mu}}
\def\Zgm{\zgm}
\def\Zz{\zzeta}

\def\gtilde{\tilde{g}_0}
\def\mtilde{\widetilde{m}_0}

\def\ms{{\rm MS}}
\def\msbar{{\rm \overline{MS\kern-0.05em}\kern0.05em}}
\def\lat{{\rm lat}}
\def\SF{\rm SF}

\newcommand{\bes}{\begin{eqnarray}}
\newcommand{\ees}{\end{eqnarray}}

\newcommand{\eq}[1]{eq.~(\ref{#1})}
\newcommand{\Eq}[1]{Eq.~(\ref{#1})}

\newcommand{\Oa}{{\rm O}(a)}
\newcommand{\fraction}[2]{{#1\over#2}}

\def\Lmax{L_{\rm max}}
\def\Lambdamsbar{\Lambda^{(0)}_{\overline{\rm MS}}}
\def\qqbar{q\overline{q}}
\def\gbarS{{\overline{g}}_{S}}
\def\gbar{{\overline{g}}}
\def\fourpi{4\pi}
\def\MSbar{\overline{\rm MS}}
\def\gbarMSbar{\overline{g}_{\overline{\rm MS}}}
\def\LambdaLat{\Lambda_{\rm L}}

\begin{titlepage}
\begin{flushright}
   ROM2F/2002/21
\end{flushright}
\vskip 1.0 cm
\begin{center}
 {\Large\bf The lattice scale at large $\beta$
            in quenched QCD\\[1.5ex]
    }
\end{center}
\vskip 0.8 cm
\begin{center}
{\large Marco Guagnelli,
        Roberto Petronzio,
        Nazario Tantalo}
\vskip 3.5ex
{\em
  Dipartimento di Fisica, Universit\`a di Roma ``Tor Vergata'' \\[1ex]
  and INFN, Sezione di Roma II \\[1ex]
  Via della Ricerca Scientifica 1, 00133 Rome, Italy
}
\end{center}
\vskip 1.0cm
\hrule
\begin{flushleft}
{\bf Abstract}
\end{flushleft}
In this paper we extend the estimate of the value of the lattice spacing
$a$ in units of the $r_0$ scale at values of the bare coupling larger than
those available. By using results from the computation of the renormalised
coupling in the Schr\"odinger functional formalism we find that from $\beta
\simeq 7$ onward the behaviour predicted by asymptotic freedom at tree loop
describes very well the data if a value of  $r_0 \Lambda$ slightly lower
then the latest available is used. We also show that, by sticking to the
current value, an effective four loop term can describe as well the data.
The systematic relative error on the lattice spacing induced by the choice
of the procedure is between 1\% and 3\% for $6.92<\beta<8.5$. \vskip
0.5truecm \hrule \vskip 1.0cm
%%%\begin{center}
%%%{September 2002}
%%%\end{center}
\vfill
\eject
\vfill
\eject
\end{titlepage}

\section{Introduction}

In order to obtain physical predictions from a lattice QCD simulation, one
has to spend as many ``experimental'' input as are the free parameters of
the theory. This situation is not peculiar of lattice regularisation, but
is a common fact to every regularisation/renormalisation procedure. In the
pure Yang--Mills case the only free parameter of the theory is the bare
coupling $g_0$, and the only dimensionful quantity in a simulation, in the
idealized case of a lattice of infinite extent, is the lattice spacing $a$.
Asymptotic freedom in the high energy regime predicts the functional
dependence of $a$ from $g_0$ once $\Lambda_L$ is known. If the value of the
bare coupling gets too large to rely on a perturbative expansion, one has
to resort to a non--perturbative definition of the physical scale of the
simulation.

One way to perform this task, in the case of a lattice extension large
enough to safely account the dynamics of the light quarks, is to use the
experimental inputs coming from the spectroscopy of the strange mesons, as
explained in refs.~\cite{Guptaseisa,Becirevicottoyg,Alltonseiyv}. Another
way to set the scale, widely used since its proposal in~\cite{Sommertrece},
is based on the calculation of the force between static colour sources; the
physical input in this case is a length scale, $r_0 \simeq 0.5\;\fm$.

In~\cite{Guagnelliottoud} and then in \cite{Neccoduemilaunoxg} the quantity
$a/r_0$ has been computed, in the quenched QCD framework, for a range of
$\beta \equiv 6/g_0^2$ going from $\beta_{\rm min} = 5.7$ to $\beta_{\rm
max} = 6.92$ and a parametric description of $\ln(a/r_0)$ as a function of
$g_0^2$ is provided that describes the data with an accuracy better than
$1\%$ in the whole range. In the quenched approximation the absolute value
of the scale is affected by a systematic error, induced by the quenching,
of the order of $10\%$; nevertheless the precise result of
ref.~\cite{Neccoduemilaunoxg} has been widely used in literature, at least
to set the ratio between different scales and to perform continuum
extrapolations. The use of the parametrization of
ref.~\cite{Neccoduemilaunoxg} outside its validity range ($5.7 \leq \beta
\le 6.92$) is highly questionable; on the other hand, one can easily
encounter situations (see for example ref.~\cite{Guagnelliduemiladuejd})
where a precise knowledge of $a/r_0$ for values of $\beta$ greater than
$6.92$ is required.

In this paper we will show how to obtain an accurate estimate of $a/r_0$
for $\beta > 6.92$, using existing numerical simulations and the behaviour
expected from asymptotic freedom: we use the non--perturbative results
coming from the computation of the Schr\"odinger functional renormalised
coupling~ \cite{Luschertregh,Capitaniottomq} and the parametrization
dictated by Renormalisation Group improved perturbation theory.

\section{Non--perturbative approach: the running coupling}

In a series of publications (see for example
refs.~\cite{Luscherduezx,Luschertregh,Capitaniottomq,Bodeduemilaunojv}) it
has been shown how to define and compute a renormalized coupling, whose
running with the energy scale can be followed, in the continuum limit, over
a very large range, making able to connect the high energy perturbative
region with the low energy non--perturbative one.

The calculations has been done using a recursive finite--size technique in
the Schr\"odinger functional framework where the energy scale is given by
$\mu = 1/L$. The size of $\Lmax$, defined implicitly by the relation
$\overline{g}^2_{\rm SF}(\Lmax) = 3.48$, has been computed in
ref.~\cite{Neccoduemilaunoxg} in terms of $r_0$, with the result
\begin{equation}
\Lmax / r_0 = 0.738(16) \;; \label{eq:lmaxr0}
\end{equation}
in the same paper, combining this result with those of
ref.~\cite{Capitaniottomq},  the authors quote
\begin{equation}
\Lambdamsbar = 0.586(48) / r_0 \;, \label{eq:lambdaMSr0}
\end{equation}
where the superscript $(0)$ remind us that we are dealing with the quenched
approximation. Equation~(\ref{eq:lmaxr0}), together with the tab.~(6) of
ref.~\cite{Capitaniottomq}, will be the handles to extend our
non--perturbative knowledge of $a/r_0$ in a wider range of $\beta$ values.

We start by considering a length scale $L_u$, implicitly defined by the
relation $\gbar^2(L_u) = u$: by construction, this scale is smaller than
$\Lmax$ by a known factor $s_u$. We can use this fact in order to express
the scale $L_u$ in terms of $r_0$:
\begin{equation}
{L_u \over r_0} \equiv x \;; \label{eq:defx}
\end{equation}
if we consider a particular discretisation of $L_u$, i.e. $L_u/a = N$, we
can write
\begin{equation}
a/r_0 = x/N \;. \label{eq:master}
\end{equation}

From table~(6) of ref.~\cite{Capitaniottomq} we can read off a whole bunch
of pairs $(L_u/a, \beta)$ and, using eq.~(\ref{eq:master}), we can  compute
$a/r_0$ corresponding to all these values of $\beta$.

Given the finite extent used in these calculations we must, of course,
assume that lattice artifacts do not play a prominent r\^ole. Whether this
is a good assumption can be checked by using the $\beta$--value
corresponding to the coupling $u =2.77$ in tab.~(6) of
ref.~\cite{Capitaniottomq}, that lies within the range in which $a/r_0$ has
been directly computed. In this case the scale factor $s_u$ is almost
exactly equal to $3/2$.

The results can be seen in table~(\ref{tab:check}): three of the values of
$a/r_0$ computed through eq.~(\ref{eq:master}), including the points
computed at $u=3.48$, are at $\beta < 6.92$, and  nicely fall, within the
error, which comes entirely from the error on $\Lmax/r_0$, over the
parametrization of ref.~\cite{Neccoduemilaunoxg}, even if the points
correspond to lattices with a relative poor discretisation, $N=8$ and
$N=12$.

Indeed, the results of ref.~\cite{Capitaniottomq} for the continuum limit
of the step scaling function also indicate that lattice artifacts are very
small already at $N=8$ and $N=12$, strongly supporting our assumption.

\begin{table}[htb]
\vskip 0.5truecm \centering
\begin{tabular} {c c c c c}
\hline \hline
$\beta$  &   eq.(2.6) ref.[6]   &  eq. (2.4)     &   $u$    &   $L_u/a$   \\
\hline
$6.4527$ &   $-2.345$           &  $-2.383(22)$  &   $3.48$ &   $8$       \\
$6.7750$ &   $-2.758$           &  $-2.790(22)$  &   $3.48$ &   $12$      \\
$6.7860$ &   $-2.773$           &  $-2.789(22)$  &   $2.77$ &   $8$       \\
\hline \hline
\end{tabular}
\caption[Table]{\footnotesize Check of eq. (2.4) against non--perturbative
result of ref. [6]. The accuracy on the parametrization of ref. [6], in
this $\beta$ range, is about 1\%. The error on the results of eq. (2.4)
comes entirely from the error on $\Lmax/r_0$. } \label{tab:check}
\end{table}

\section{Perturbative approach}

In ref.~\cite{Neccoduemilaunogh} it is argued that the three--loop
perturbative expansion in the $\qqbar$--scheme may be trusted up to
$\alpha_{\qqbar}\simeq 0.3$. We test how perturbation theory can help us in
safely extending the region where $a/r_0$ has to be considered as known
with good precision, contrary to the general belief that lattice bare
perturbation theory is well-behaving only in the extremely huge cutoff
region.

We recall some well known results. In a generic renormalisation scheme,
$S$, the perturbative expansion of the $\beta$--function reads:
\begin{equation}
\mu {{\rm d}\over{\rm d}\mu} \gbarS \equiv \beta(\gbarS) = -\gbarS^3\{ b_0
+ b_1 \gbarS^2 + b_2^{(S)}\gbarS^4 + ...\} \label{eq:betafun}
\end{equation}
where $b_0$ and $b_1$ are the well--known universal terms,
\begin{equation}
b_0 = {1\over{(\fourpi)^2}}(11 - {2\over 3} N_f),\quad\quad b_1 =
{1\over{(\fourpi)^4}}(102 - {38\over 3} N_f) \label{eq:universalterms}
\end{equation}
while, from $b_2^{(S)}$ on, the coefficients of the $\beta$--function start
to be scheme--dependent. The $b_2^{(S)}$ coefficient is known in the
$\MSbar$ scheme~\cite{Tarasovau}:
\begin{equation}
b_2^{(\MSbar)} = {1\over{(\fourpi)^6}}({77139\over 54} - {5033\over 18} N_f
+ {325\over 54} N_f^2 ). \label{eq:b2msbar}
\end{equation}
Since the relation between $\gbarMSbar$ and the bare lattice coupling $g_0$
is known up to two loops \cite{martinpeterone,martinpetertwo}, some
straightforward algebra leads us to the $b^{L}_2$ coefficient of the
lattice $\beta$--function, that we quote in the $N_f=0$ case only:
\begin{equation}
b^{L}_2 = b^{\rm\overline{MS}}_2 + b_1 d_1(1) / ( 4\pi ) +
(d_1^2(1)-d_2(1))b_0 / (4\pi)^2 = -0.0015998323314(3) \label{eq:b2lat}
\end{equation}
where $d_1$ and $d_2$ can be found in ref. \cite{martinpeterone}.

In order to integrate the renormalisation group equations, we need the
integration constant, usually defined trough the renormalisation QCD scale
$\Lambda_S$, which can be related to the running coupling $\gbarS(\mu)$
through the exact relation
\begin{equation} \Lambda_S =
\mu(b_0\gbarS^2)^{-b_1/(2b_0^2)}{\rm e}^{-1/(2b_0\gbarS^2)}
\exp\Bigl\{-\int_0^{\gbarS} {\rm d}x\bigl[{1\over\beta(x)} + {1\over{b_0
x^3}} - {b_1\over{b_0^2x}}\bigr]\Bigl\}. \label{eq:deflambda}
\end{equation}
where the subscript $S$ indicates a generic scheme. The previous expression
obviously becomes a perturbative one upon the insertion of the perturbative
formula (\ref{eq:betafun}) for the $\beta$--function. In the case of
lattice bare perturbation theory, we can recast the previous equation in
the following form:
\begin{equation}
\ln(a/r_0) = -\ln(\LambdaLat r_0) - {b_1\over{2b_0^2}}\ln(b_0g_0^2) -
{1\over{2b_0g_0^2}} - I(g_0) \;, \label{eq:defaoverr0}
\end{equation}
where $I(g_0)$ is the integral appearing in the exponential of eq.
(\ref{eq:deflambda}). By using eq.~(\ref{eq:lambdaMSr0}) and the result
from ref. \cite{msbarlat}
\begin{equation}
\Lambdamsbar / \LambdaLat = 28.80934(1) \label{eq:lambdaratio}
\end{equation}
we get
\begin{equation}
\LambdaLat r_0 = 0.0203(17) \label{eq:lambdaLatr0}
\end{equation}
and, with this input, eq.~(3.6) becomes a perturbative tool to set the
scale $a$ in terms of $r_0$.

\begin{figure}[htbp]
  \begin{flushleft}
   \epsfig{figure=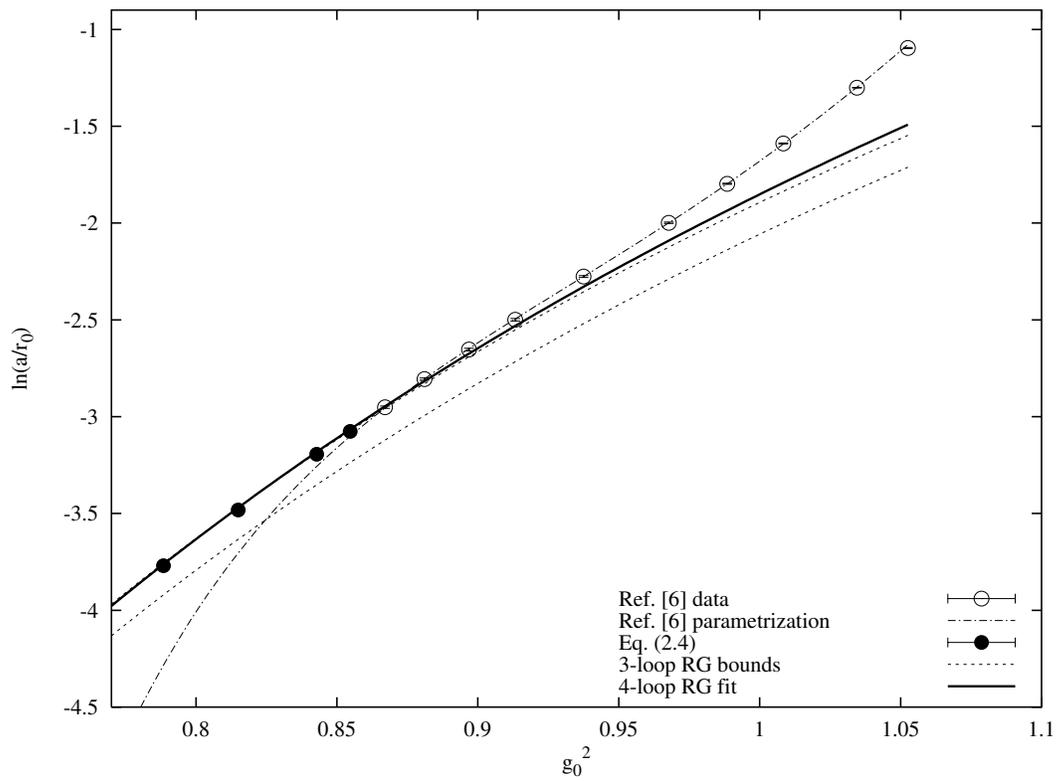, width=10.5cm, angle=270}
    \caption{\footnotesize
     $\ln(a/r_0)$ as a function of $g_0^2$.
    }
    \label{fig:res}
  \end{flushleft}
\end{figure}

As can be seen in fig. (\ref{fig:res}), the three--loops formula describes
quite well the behaviour of non--perturbative data at high $\beta$ and
nicely superimpose to the points computed through eq. (2.4) if the value of
$\Lambda$ is lowered by about $1\,\sigma$ with respect to the value quoted
in ref.~\cite{Neccoduemilaunoxg} (the upper bound of the three--loops band
in the figure corresponds to $\Lambdamsbar = 0.538/r_0$).

If we do not stick to the value $\Lambdamsbar = 0.586/r_0$ but let the
three-loops fit decide for its value, we obtain the results listed in table
\ref{tab:fit3loop}. The value of $\Lambdamsbar$ obtained through this
procedure is remarkably constant against the variation of the number of
points included in the fit. Whether this fact is really suggestive of a
lower value of the $\Lambda$ parameter in quenched approximation could be
only clarified with further non--perturbative investigations at smaller
values of the bare coupling.

Fig. (\ref{fig:res}) indicates that, at $\beta \simeq 7$, three-loops
perturbation theory is not far from an accurate description of
non--perturbative data even when we keep $\Lambdamsbar = 0.586 / r_0$. We
can thus try to determine an effective four--loops term by means of a fit.
A two parameters fit ($\Lambda$ and $b_3$) gives a value for $\Lambdamsbar
r_0$ totally compatible, in the error, with $0.586$, and this make us
confident that reducing the number of parameters by sticking to the known
value of $\Lambdamsbar r_0$ will give a reliable estimate of $b_3$.

In table (\ref{tab:fit4loopb3only}) we show the results obtained with a
one--parameter fit, namely $b_3^{\rm eff}$ itself, by keeping fixed
$\Lambdamsbar = 0.586 / r_0$. We note that absolute value of $b_3^{\rm
eff}$ agrees with an approximate geometric growth of the coefficients
$(4\pi)^{2(n+1)}b_n$.

\begin{table}[htb]
\vskip 0.5truecm
\centering
\begin{tabular} {c c c c}
\hline
\hline
$n$  &   $\Lambdamsbar r_0$   &  $\ln(a/r_0)(\beta=7.5)$ &  $\ln(a/r_0)(\beta=8.5)$  \\
\hline
$3$ &   $0.547$ &   $-3.640$  &  $-4.801$   \\
$4$ &   $0.546$ &   $-3.638$  &  $-4.800$   \\
$5$ &   $0.540$ &   $-3.627$  &  $-4.789$   \\
$6$ &   $0.535$ &   $-3.618$  &  $-4.779$   \\
\hline
\hline
\end{tabular}
\caption[Table]{\footnotesize Results for the 3--loop fit to
non--perturbative data at large $\beta$, including the last four points
obtained through eq. (2.4); $n$ is the number of points included in the fit
(most perturbative ones). } \label{tab:fit3loop}
\end{table}

\begin{table}[htb]
\vskip 0.5truecm
\centering
\begin{tabular} {c c c c}
\hline
\hline
$n$  &   $b_3^{\rm eff}$   &  $\ln(a/r_0)(\beta=7.5)$  &  $\ln(a/r_0)(\beta=8.5)$ \\
\hline
$3$ &   $-0.0022$ &   $-3.643$  &  $-4.819$   \\
$4$ &   $-0.0022$ &   $-3.643$  &  $-4.818$   \\
$5$ &   $-0.0023$ &   $-3.638$  &  $-4.815$   \\
$6$ &   $-0.0025$ &   $-3.631$  &  $-4.810$   \\
\hline
\hline
\end{tabular}
\caption[Table]{\footnotesize Results for the 4--loop fit to
non--perturbative data at large $\beta$, including the last four points
obtained through eq. (2.4), keeping fixed $\Lambdamsbar r_0 = 0.586$; $n$
is the number of points included in the fit (most perturbative ones). }
\label{tab:fit4loopb3only}
\end{table}

Summarizing, it is safe to assume that beyond $\beta = 6.92$ the quantity
$\ln(a/r_0)$ is well described by an effective four--loops perturbative
curve, if not already by the three-loops expression. We give as a final
result the value obtained with $n=6$ and $\Lambdamsbar r_0$ fixed to the
value $0.586$ ($b_3^{\rm eff} = -0.0025(3)$: the $\chi^2/{\rm n.d.f.}$ of
the fit is $0.9$), that we chose in order to obtain the lattice scale $a$
in terms of $r_0$ for $\beta > 6.92$: the fit is shown in fig.
(\ref{fig:res}). The accuracy of this description can be estimated by
considering the variation of this quantity when $\Lambdamsbar$ changes by
$1\,\sigma$. Of course the value of $b_3^{\rm eff}$ will be readjusted by
the fit. We take as a reference points the values of $\ln(a/r_0)$ at $\beta
= 7.5$ and $\beta = 8.5$, where we obtain respectively (assuming $r_0 =
0.5\,\fm$)
\begin{eqnarray}
\beta &=& 7.5,\quad\quad\quad\ln(a/r_0) = -3.63(2)\,\quad\quad a =
0.0133(3)\,
\fm,\nonumber\\
\beta &=& 8.5,\quad\quad\quad\ln(a/r_0) = -4.81(3)\,\quad\quad a =
0.0041(1)\, \fm
\end{eqnarray}
We can therefore state that, no matter which solution we adopt, lower
$\Lambda$ and 3--loops or standard $\Lambda$ and 4--loops, the uncertainty
on $a/r_0$ for $6.92 < \beta < 7.5$ can be bounded by 2\%, raising to $3\%$
at $\beta = 8.5$ and asymptotically reaching the value of the uncertainty
on $\Lambda$ of $8\%$.

\section*{Aknowledgments}
We thank R. Sommer for a critical reading of the manuscript.

\end{document}